\documentclass[conference,letterpaper]{IEEEtran}

\usepackage[utf8]{inputenc} 
\usepackage[T1]{fontenc}
\usepackage{url}
\usepackage{ifthen}
\usepackage{cite}
\usepackage[cmex10]{amsmath} 

\usepackage{amsmath}
\usepackage{amssymb} 
\usepackage{comment}

\usepackage{color, colordvi}
\usepackage{float} 

\def\theenumi{\alph{enumi}}

\def\theenumii{\roman{enumii}}

\def\Reals{\mathbb R}             

\def\L1{{\mathcal L}^{1}(\Reals)} 
\def\L2{{\mathcal L}^{2}(\Reals)} 
\def\L2T{{\mathcal L}^{2}\left( (T_0, T_1) \right)} 
\def\Lp2{{\mathcal L}^{2}(\Reals, p(\cdot))} 

\def\P{\text{P}}                                  
\newcommand{\bP}{{\bf P}}

\def\P_e{\bP_e}                                   
\newcommand{\E}[1]{\mathsf{E}\left[{#1}\right]}   

\newcommand{\eqdef}{\; \hat{=} \;}

\def\N0{{\mathcal N}_{0}}                 

\def\sgn{\mathsf{sgn}}  
 
\newcommand{\lmvue}[1]{\lambda_{\text{MVUE}}}

\newcommand{\vect}[1]{\mathbf{#1}}   

\newcommand{\vG}{\vect{G}}

\makeatletter 
\@ifundefined{theorem}{%
\newtheorem{theorem}{Theorem}
\newtheorem{lemma}{Lemma}

\newtheorem{corollary}{Corollary}

}{}
\makeatother

\definecolor{Light}{gray}{.93}

\def\SaS{\textrm{S}$\alpha$\textrm{S} }

\begin{document}
\title{On Linear Estimators for some Stable Vectors} 


\author{%
  \IEEEauthorblockN{Rayan Chouity, Charbel Hannoun, Jihad Fahs and Ibrahim Abou-Faycal}
  \IEEEauthorblockA{Department of Electrical and Computer Engineering\\ 
                    American University of Beirut\\
                    Beirut, Lebanon\\
                    Email: \{rac45,cnh10,jihad.fahs,ibrahim.abou-faycal\}@aub.edu.lb}
}

\maketitle

%
\begin{abstract}
We consider the estimation problem for jointly stable random variables. Under two specific dependency models: a linear transformation of two independent stable variables and a sub-Gaussian symmetric $\alpha$-stable (\SaS\!) vector, we show that the conditional mean estimator is linear in both cases. Moreover, we find dispersion optimal linear estimators. Interestingly, for the sub-Gaussian \SaS vector, both estimators are identical generalizing the well-known Gaussian result of the conditional mean being the best linear minimum-mean square estimator.  
\end{abstract}

\section{Related Work}

The problem of estimating a random or deterministic quantity after observing another related variable is a classical problem in estimation theory. One of the most prevalent formulation of the estimation problem is a Bayesian one under which the conditional expectation plays a central role as it is optimal under mean-squared error loss. Beyond that, the conditional mean was shown in \cite{banerjee2005bregman} to be, under mild conditions, Bayes-optimal for a broad class of loss functions, namely Bregman divergences. Additionally, in additive Gaussian noise models, the conditional mean admits derivative relationships that characterize higher-order conditional quantities and facilitate conditional-moment estimation \cite{meta_derivative_identity_gaussian}.

A complementary line of research asks when are Bayes-optimal estimators linear (or affine) functions. In additive Gaussian noise settings, the authors in~\cite{L1_optimality_linear_estimators} showed that under $L_p$ losses where $p \in [1, 2]$, Bayes-optimal estimators are linear if and only if the prior distribution is Gaussian. Whenever $p > 2$, it is shown in~\cite{L1_optimality_linear_estimators} that there exist infinitely many priors that yield linear optimal estimators. These results were extended to multivariate settings in~\cite{multivariate_prior}. The framework of finding priors that induce linear optimal estimators has also been considered beyond the Gaussian noise model as in~\cite{L1_poisson}, where $L_1$ linear optimal estimators where studied under Poisson loss.

Noise models that do not fall under the exponential type~\cite{LUO2013650} are still underexplored in the literature. This is the case for example, for power-law heavy-tailed models for which the moment generating function does not exist. A canonical class is given by $\alpha$-stable laws --a natural heavy-tailed counterpart of the Gaussian law by virtue of the Generalized Central Limit Theorem (GCLT)~\cite{kolmo1968}, that arise in interference models~\cite{hughes2000alpha,win2009mathematical} and molecular communications~\cite{Farsa15}, for instance. Such distributions are described (up to location) by a spectral measure on the unit sphere~\cite{sam1994}, and conditional moments are governed by integrability properties of this measure. In particular, the authors in~\cite{cioczek1994_dependence_spectral} related the spectral measure to the existence of conditional moments, and derived in~\cite{cioczek1995_necessary} necessary conditions. Sufficient conditions for the existence of the conditional moments were established~\cite{taqqu_chapter_conditional} for specific parameter ranges.

Although several cost metrics can be imposed on the quality of the estimation strategy for heavy-tailed variables such as~\cite{Shao1993,gonza2006,FAF2018,verdu2023}, a primary objective of this work is to analyze the conditional mean estimator for jointly Symmetric $\alpha$-Stable (\SaS\!\!\!) variables\footnote{A symmetric $\alpha$-stable law will be denoted by $\mathcal{S}(\alpha, \gamma)$ where $0 < \alpha < 2$ is the stability parameter and $\gamma > 0$ the dispersion.} --having infinite second moments and possibly infinite means, under two specific dependency schemes: A linear transformation of independent ones and the sub-Gaussian model. We note that these two schemes cover widely used dependency relationships such as the additive noise model and  elliptical vectors.   

Our main contributions are as follows: We show the existence and linearity of the conditional expectation estimator for the considered $\alpha$-stable settings. Moreover, we find dispersion-optimal linear estimators for both setups. Our results reveal that the conditional expectation and the minimum dispersion linear estimators are different for the linear mix case and a fortiori for the special case of the additive channel model. However, both linear estimators are identical for the \SaS sub-Gaussian dependency model recovering the well-known Gaussian result of the conditional expectation being the best linear mean-square error estimator. This suggests that the natural heavy-tailed extension of the Gaussian framework is the elliptical sub-Gaussian \SaS dependency model and not the linear mix of independent \SaS variables.  


\section{Models}

In this work we study the problem of estimating an absolutely continuous random quantity $X$ based on observing a related random quantity $Y$ whenever the vector 
\begin{equation*}
\begin{pmatrix}
    X \\ Y
\end{pmatrix}
\end{equation*}
is a stable vector. We restrict the analysis to scenarios where both $X$ and $Y$ have symmetric Probability Density Functions (PDF). In the Gaussian case, the (Gaussian) vector at hand can both
\begin{itemize}
    \item be written as a linear function of independent Gaussian variables.
     \item be seen as one with a PDF the contours of which are ellipsoids.
\end{itemize}

For $\alpha$-stable vectors, the two views are {\em not\/} equivalent and actually form disjoint sets of stable vectors. In this study we treat both models and we label them as:

\begin{itemize}
\item {\bf linear-mix} where the vector is
    \begin{equation}
    \begin{pmatrix}
    X \\ Y
    \end{pmatrix} =  \begin{pmatrix}
    a_{11} & a_{12} \\ a_{21} & a_{22}
    \end{pmatrix}
    \begin{pmatrix}
    Z_1 \\ Z_2
    \end{pmatrix},
    \label{eq:linmixgen}
    \end{equation}
    where $Z_1$ and $Z_2$ are independent \SaS variables with stability parameter $\alpha$ and dispersion (scale) parameters $\gamma_{Z_1}$ and $\gamma_{Z_2}$ respectively.

    Before proceeding, note that whenever a column in the matrix is zero, one variable is a linear scale of the other. Additionally, whenever one of terms in the second row is zero, the variable $X$ has an independent $\alpha$-stable component from $Y$. We eliminate both scenarios from our analysis as in the first case the problem is trivial and in the second an observation $Y$ changes essentially the location of an $\alpha$-stable variable. 

    
    In summary, We limit our analysis to setups where
    \begin{equation}
        a_{21} \neq 0 \qquad \& \qquad  a_{22}\neq 0
        \label{eq:Cond}
    \end{equation} 
    and where the matrix is full-rank.  
    
    \item {\bf sub-Gaussian} where the vector is
    \begin{equation}
    \begin{pmatrix}
    X \\ Y
    \end{pmatrix} = \sqrt{A} \, \vG \eqdef 
    \sqrt{A} \begin{pmatrix}
    G_1 \\ G_2
    \end{pmatrix},
    \label{eq:subGauss}
    \end{equation}
    where $\vG$ is a Gaussian vector that is zero-mean and with covariance matrix $\Sigma$, and $A \sim \mathcal{S}\left(\frac{\alpha}{2},1,\cos^{\frac{2}{\alpha}}\left(\frac{\pi \alpha}{4}\right),0\right)$ is a non-negative $\alpha/2$-stable random variable, $0 < \alpha < 2$ with dispersion equal to $\cos^{\frac{2}{\alpha}}\left(\frac{\pi \alpha}{4}\right)$~\cite{sam1994}.
\end{itemize}

Interestingly, for the model in~\eqref{eq:linmixgen}, and contrary to the Gaussian case, the vector $(X,Y)^{\top}$ cannot be independent (Darmois–Skitovich theorem~\cite{darmois1953,skito1953}) unless the matrix is trivially the identity matrix, a case that we discarded here. We note that for the special case of $a_{11} = 1, a_{12} = 0$,
the model is equivalent to an additive noise model $Y = a_{21} X + a_{22} Z_2$. The second model~\eqref{eq:subGauss} covers the special isotropic case, which is a dependent vector unlike the isotropic Gaussian vector (Herschel-Maxwell's theorem~\cite[Thm. 0.0.1]{bryc2012normal}).

\subsection{Criteria}

As in any estimation framework, the "quality" of the estimate $\hat{X}(Y)$ of $X$ --based on observing $Y$-- is quantified through a fidelity criterion $c(x, \hat{x})$, and one is naturally interested in finding an "optimal" estimator $\hat{X}(\cdot)$ if such one exists. The quality metric $c(\cdot, \cdot)$ is a deterministic scalar that is chosen with an application in mind: a particular metric may seem more suitable than another for a specific application at hand. The merits of a particular metric versus another is not the subject of this work, however, we argue some choices are more sensible than others given that the data at hand is heavy-tailed.

The most common criteria used in the literature are almost all of the form:
\begin{equation*}
    \E{c(X, f(Y))},
\end{equation*}
where the expectation is naturally over the variables $X$ and $Y$.
\begin{itemize}
    \item[$\bullet$] If $c(a, b) = (a - b)^2$, the Bayesian Least Squares Estimator (BLSE) $\hat{X}(Y) = \E{X | Y}$ is found optimal through the use of iterated expectations:
    \begin{equation*}
        \E{(X - \hat{X}(Y))^2} = \E{\E{(X - \hat{X}(Y))^2 | Y }}.
    \end{equation*}
    
    Intuitively, one can easily support the simple choice --given an observation $Y$-- of the average value of $X$.

    \item[$\bullet$] If $c(a, b) = |a - b|$, the Minimum Absolute Estimator (MAE) $\hat{X}(Y) = \text{median of } X$ given $Y$ is optimal. It is interesting to note that the optimal estimator in this case is also --as above, a first-order location quantity of the conditional probability law of $X$ given $Y$.

    \item[$\bullet$] Whenever 
    \[
    c(a, b) = \left\{ \begin{array}{ll}
    1 & \text{ if } |a - b | \geq \epsilon \\
    0 & \text{ otherwise, }
    \end{array} \right.
    \]
    and as $\epsilon$ goes to zero, the optimal estimator is the Maximum-Likelihood (ML) estimator. This is also a first-order location quantity (the mode) of the conditional probability law of $X$ given $Y$. 

    Perhaps the ML estimator can be best intuitively championed, as it is simply the most likely value of $X$ given a specific observation.
\end{itemize}
Interestingly enough, all of the above estimators coincide whenever the conditional density of $(X \mid Y=y)$ is unimodal and symmetric, as is the case of jointly Gaussian variables, for example. 

The virtues of linear estimators have also been argued in the literature at length, including in almost all textbook materials. 

In accordance with the above, we will pay special attention to linear estimators for our problem and highlight possible criteria, their advantages, and their shortcomings.

\section{Linear-Mix}


The conditional probability law of $X$ given the observation $y$,  $p_{X|Y}(x, y)$ can be determined by Bayes' rule:
\begin{equation*}
p_{X|Y} (x|y) = \frac{p_{Y|X}(y|x) \, p_X(x)}{p_Y(y)}.
\end{equation*}

Alternatively, one can work ``in the Fourier domain" or with characteristic functions since the vast majority of \SaS random variables can only be described by their characteristic functions which we denote by $\phi(t)$. Indeed, if $U \sim \mathcal{S}(\alpha, \gamma_U)$ then $\phi_U(t) = \exp \left( -\gamma_U^\alpha |t|^\alpha \right)$.

\subsection{A Linear Conditional Expectation Estimator}
\label{sec:condmean}
Next, we state and prove our first result. 

\begin{theorem}[Linear Mixture Conditional Expectation]
\label{th:linearcondexp}
Under the linear-mix model given in equation~\eqref{eq:linmixgen} and subject to conditions~\eqref{eq:Cond}, for a given observation $Y=y$, the conditional expectation $\E{X \mid Y = y}$ of the \SaS variable $X$ exists, is linear in $y$ and is given by
\begin{align*}
& \E{X|Y=y} \\
&= \frac{(|a_{21}|\gamma_{Z_1})^\alpha \left[\frac{a_{11}}{a_{21}} \right] + (|a_{22}| \gamma_{Z_2})^\alpha \left[ \frac{a_{12}}{a_{22}}\right] }{(|a_{21}|\gamma_{Z_1})^\alpha + (|a_{22}| \gamma_{Z_2})^\alpha)} \, y.
\end{align*}
In particular, whenever $a_{11} = 1, a_{12} = 0$, $Y = a_{21}X + a_{22}{Z_2}$ and
\begin{equation*}
\E{X|Y=y} = \frac{\gamma_{X}^\alpha |a_{21}|^{\alpha}}{\gamma_{X}^\alpha|a_{21}|^\alpha+\gamma_{Z_2}^\alpha|a_{22}|^\alpha} \, \frac{y}{a_{21}}.
\end{equation*}
\end{theorem}

\begin{IEEEproof}
Let $Z_1$ and $Z_2$ be independent \SaS variables with respective scales $\gamma_{Z_1}$ and $\gamma_{Z_2}$. 
According to the model defined by equation~\eqref{eq:linmixgen}, both $X$ and $Y$ are linear combinations of independent \SaS random variables. By the stability property $X \sim \mathcal{S}(\alpha, \gamma_{X})$, and $Y \sim \mathcal{S}(\alpha, \gamma_{Y})$
with scale parameters $\gamma_{X}^\alpha
=|a_{11}|^\alpha \gamma_{Z_1}^\alpha+|a_{12}|^\alpha \gamma_{Z_2}^\alpha$ and $\gamma_{Y}^\alpha
=|a_{21}|^\alpha \gamma_{Z_1}^\alpha+|a_{22}|^\alpha \gamma_{Z_2}^\alpha$ respectively. Furthermore, $X$ and $Y$ are characterized by the joint probability density $p_{X,Y}(x,y)$ and characteristic function
\begin{align}
\phi_{X,Y}(t,s)
&=\E{e^{jtX}e^{jsY}} \nonumber\\
&=\E{e^{jt(a_{11}Z_1+a_{12}Z_2) + js(a_{21}Z_1+a_{22}Z_2)}}\nonumber\\
&=\E{e^{j(a_{11}t+a_{21}s)Z_1}}\E{e^{j(a_{12}t+a_{22}s)Z_2}} \label{eq:indZ12}\\
&=\phi_{Z_1}(a_{11}t+a_{21}s)\phi_{Z_2}(a_{12}t+a_{22}s)\nonumber\\
&= e^{ -\gamma_{Z_1}^\alpha|a_{11}t+a_{21}s|^\alpha-\gamma_{Z_2}^\alpha|a_{12}t+a_{22}s|^\alpha}, \label{eq:jointcharadef} \\
& = e^{-\gamma_1^\alpha|s-k_1t|^\alpha-\gamma_2^\alpha|s-k_2t|^\alpha}
\end{align}
where $k_1 \triangleq -\frac{a_{11}}{a_{21}}$, $k_2 \triangleq -\frac{a_{12}}{a_{22}}$,  $\gamma_1 \triangleq \gamma_{Z_1}|a_{21}|$ and $\gamma_2 \triangleq \gamma_{Z_2}|a_{22}|$. Note that the independence of $Z_1$ and $Z_2$ allows us to write equation~\eqref{eq:indZ12}. 

Let $\phi_{X|Y=y}(t,y)$ be the characteristic function of variable $X$ conditioned on the event that $Y = y$. Using the integral form, $\phi_{X,Y}(t,s)$ can be written as the Fourier transform (up to a sign change) of the function of $y$: $p_Y(y)\phi_{X|Y}(t,y)$ 
    \begin{align*}
    \phi_{X,Y}(t,s)
    &=\int_\mathbb{R}\int_\mathbb{R} p_{X,Y}(x,y)e^{jtx+jsy}dxdy\\
    &=\int_\mathbb{R}\int_\mathbb{R} p_Y(y)p_{X|Y}(x|y)e^{jtx+jsy}dxdy\\
    &=\int_\mathbb{R}p_Y(y)\left(\int_\mathbb{R} p_{X|Y}(x|y)e^{jtx}dx\right)e^{jsy}dy\\
    &=\int_\mathbb{R}p_{Y}(y)\phi_{X|Y}(t,y)e^{jsy}dy.
    \end{align*}
    Therefore, applying an inverse Fourier transform we get
    \[\phi_{X|Y}(t,y) = \frac{1}{2\pi}\frac{I(t,y)}{p_{Y}(y)},\]
    where 
    \[I(t,y)=\int_{-\infty}^{+\infty}\phi_{X,Y}(t,s)e^{-jsy}ds.\]
The conditional expectation of $X$ given an observation $Y = y$ is equal to
\begin{align}
    \E{X|Y=y} &= \int_{-\infty}^{+\infty} x p_{X|Y}(x|y)\,dx \nonumber\\
    &= \frac{1}{j} \frac{\partial}{\partial t}\left[\int_{-\infty}^{+\infty}p_{X|Y}(x|y)e^{jtx} \, dx \right]_{t=0} \nonumber\\
    &= \frac{1}{j} \frac{\partial\phi_{X|Y}}{\partial t}(0,y) = \frac{1}{2\pi j \, p_{Y}(y)}\frac{\partial I}{\partial t}(0,y). \label{eq:condmean}
\end{align}

Whenever the derivative of $I(\cdot,y)$ exists and is finite for all $y \in \mathbb{R}$, $\E{X \mid Y = y}$ exists --a fact that we will show next. We compute next this derivative $\frac{\partial}{\partial t} I(t,y)$  assuming $k_1 t \leq k_2 t$. Let
\[
I(t,y)=f(t,y)+m(t,y)+l(t,y),
\]
where
\begin{align*}
    f(t,y)&=\int_{-\infty}^{k_1t}e^{-\gamma_1^\alpha(k_1t-s)^\alpha-\gamma_2^\alpha(k_2t-s)^\alpha - jsy}ds\\
    m(t,y)&=\int_{k_1t}^{k_2t}e^{-\gamma_1^\alpha(s-k_1t)^\alpha-\gamma_2^\alpha(k_2t-s)^\alpha - jsy}ds\\
    l(t,y)&=\int_{k_2t}^{+\infty}e^{-\gamma_1^\alpha(s-k_1t)^\alpha-\gamma_2^\alpha(s-k_2t)^\alpha - jsy}ds
\end{align*}
Applying the Leibniz integration rule, and with 
\begin{equation}
    \label{eq:intg}
    g(t,s,y) \triangleq \phi_{X,Y}(t,s)e^{-jsy}
\end{equation}
and 
\begin{align}
\label{eq:inth}
h(t,s)&\triangleq\alpha\gamma_1^\alpha k_1\sgn(s-k_1t)|s-k_1t|^{\alpha-1}\nonumber\\&+\alpha\gamma_2^\alpha k_2\sgn(s-k_2t)|s-k_2t|^{\alpha-1},
\end{align}
we write
\begin{align*}
& \frac{\partial f}{\partial t}(t,y)=
    k_1\,g(t,k_1t,y) + \int_{-\infty}^{k_1t} h(t,s) g(t,s,y) ds\\
& \frac{\partial m}{\partial t}(t,y)
    = \Big[ k_2g(t,k_2t,y) -k_1 g(t,k_1t,y) \Big. \\ 
    & \qquad \qquad + \left. \int_{k_1t}^{k_2t} h(t,s) g(t,s,y) ds \right] \\
& \frac{\partial l}{\partial t}(t,y)
   = -k_2 g(t,k_2t,y) + \int_{k_2t}^{+\infty} h(t,s) g(t,s,y) ds
\end{align*}
Since $g(0,0,y) = 1$, we notice that
\begin{align*}
&\frac{\partial I}{\partial t}(0,y) \\
&= \int_{-\infty}^{+\infty} h(0,s)g(0,s,y)\,ds\\
&= \alpha (\gamma_1^{\alpha}k_1 + \gamma_2^{\alpha}k_2)\!\!\int_{-\infty}^{+\infty} \!\! \sgn(s)|s|^{\alpha-1}e^{-(\gamma_1^\alpha + \gamma_2^\alpha)|s|^{\alpha}}e^{-jsy}\,ds,
\end{align*}
which represents the Fourier transform of an $L^{1}$ function, for $0 < \alpha < 2$, thus implying that $\frac{\partial I}{\partial{t}}(0,y)$ exists. 
Defining $\tilde{\gamma}^\alpha \triangleq k_1\gamma_{1}^\alpha + k_2\gamma_{2}^\alpha$, we therefore have 
\begin{equation}
\frac{\partial I}{\partial t}(0,y) = 2j \tilde{\gamma}^\alpha \int_{0}^{+\infty}(-\alpha s^{\alpha-1} e^{-\gamma_{Y}^\alpha|s|^\alpha}) \sin(sy)\,ds, \label{eq:byparts}
\end{equation}
where $\gamma_{Y}^\alpha = \gamma_1^{\alpha} + \gamma_2^{\alpha} = |a_{21}|^\alpha \gamma_{Z_1}^\alpha+|a_{22}|^\alpha \gamma_{Z_2}^\alpha$ as previously defined and where we used the fact that $\sgn(s)|s|^{\alpha-1}e^{-(\gamma_1^\alpha + \gamma_2^\alpha)|s|^{\alpha}}$ is an odd function. 
            
Integrating equation~\eqref{eq:byparts} by parts, and noticing that 
\[\frac{1}{2\pi}\int_{0}^{+\infty}e^{-\gamma_{Y}^\alpha|s|^\alpha}\cos(sy)\,ds=\frac{1}{2}p_{Y}(y),\]
we get the following:
\begin{align*}
    \frac{\partial I}{\partial t}(0,y)
    & = \frac{2 j \tilde{\gamma}^\alpha}{\gamma_{Y}^\alpha} \left[ \left[ e^{-\gamma_{Y}^\alpha |s|^\alpha} \sin(sy) \right]_{s=0}^{+\infty} \right.\\
    &\qquad\qquad\quad\left.-y\int_{0}^{+\infty}e^{-\gamma_{Y}^\alpha|s|^\alpha}\cos(sy)ds\right]\\
    &=-\frac{2\pi j \tilde{\gamma}^\alpha}{\gamma_{Y}^\alpha} \,y \,p_{Y}(y).
\end{align*}

Therefore equation~\eqref{eq:condmean} becomes
\begin{align}
    &\E{X|Y=y} = -\frac{\tilde{\gamma}^\alpha}{\gamma_Y^\alpha}y\nonumber \\
    &=-\frac{k_1\gamma_{Z_1}^\alpha|a_{21}|^\alpha+k_2\gamma_{Z_2}^\alpha|a_{22}|^\alpha}{\gamma_{Z_1}^\alpha|a_{21}|^\alpha+\gamma_{Z_2}^\alpha|a_{22}|^\alpha}\,y\nonumber\\
    &=\frac{\gamma_{Z_1}^\alpha |a_{21}|^{\alpha} \left[ \frac{a_{11}}{a_{21}} \right] + \gamma_{Z_2}^\alpha |a_{22}|^{\alpha} \left[ \frac{a_{12}}{a_{22}}\right] }{\gamma_{Z_1}^\alpha|a_{21}|^\alpha+\gamma_{Z_2}^\alpha|a_{22}|^\alpha} \, y \label{eq:condexpct}
\end{align}

Now consider the scenario $k_1t > k_2t$. Identical results are obtained, by simply exchanging the subscripts $1$ and $2$:
\begin{align*}
    f(t,y)&=\int_{-\infty}^{k_2t}e^{-\gamma_1^\alpha(k_1t-s)^\alpha-\gamma_2^\alpha(k_2t-s)^\alpha - jsy}ds\\
    m(t,y)&=\int_{k_2t}^{k_1t}e^{-\gamma_1^\alpha(k_1t-s)^\alpha-\gamma_2^\alpha(s-k_2t)^\alpha - jsy}ds\\
    l(t,y)&=\int_{k_1t}^{+\infty}e^{-\gamma_1^\alpha(s-k_1t)^\alpha-\gamma_2^\alpha(s-k_2t)^\alpha - jsy}ds
\end{align*}
Applying the Leibniz integration rule, and with $g(t,s,y)$ and $h(t,s)$ as defined in \eqref{eq:intg} and \eqref{eq:inth}, we write
\begin{align*}
    &\frac{\partial f}{\partial t}(t,y)=
    k_2 \, g(t,k_2t,y) + \int_{-\infty}^{k_2t}h(t,s)g(t,s,y)ds\\
    &\frac{\partial m}{\partial t}(t,y)
    =\left[k_1g(t,k_1t,y)-k_2g(t,k_2t,y)\right.\\
    & \qquad \qquad + \left. \int_{k_2t}^{k_1t} h(t,s) g(t,s,y) ds\right]\\
    &\frac{\partial l}{\partial t}(t,y)
    =-k_1g(t,k_1t,y)+\int_{k_1t}^{+\infty}h(t,s)g(t,s,y)ds
\end{align*}
Notice that \(\frac{\partial I}{\partial t}(0,y)\)
exists and is the same as in Equation~\eqref{eq:byparts} for case $k_1t\leq k_2t$. This implies the same resulting expression as in \eqref{eq:condexpct}

\noindent
\paragraph*{Special Case} In the specific case of the additive noise model where the observation $Y$ is 
\begin{equation*}
Y = a_{21} X + a_{22} Z_2,
\end{equation*}
the observation is a \SaS with scale parameter
\(
\gamma_Y = \left( |a_{21}|^\alpha \gamma_{Z_1}^\alpha + |a_{22}|^{\alpha} \gamma_{Z_2}^\alpha \right)^{\frac{1}{\alpha}} .
\)
In this case, equation (\ref{eq:condexpct}) simplifies to
\begin{equation}
\label{eq:condlinmix}
\E{X|Y=y} = \frac{\gamma_{X}^\alpha |a_{21}|^{\alpha}}{\gamma_{X}^\alpha|a_{21}|^\alpha+\gamma_{Z_2}^\alpha|a_{22}|^\alpha}\, \frac{y}{a_{21}}.
\end{equation}
\end{IEEEproof}

\subsection{The Dispersion-Optimal Linear Estimator}
\label{sec:optim}
The result of Theorem~\ref{th:linearcondexp} establishes the linearity of the condition expectation --optimal under certain ``quality" criterion such as a suitably chosen Bregman divergence whenever the vector is stable. One can notice that a linear estimator leads to a stable error. This observation raises the question of what is the best estimator among those linear ones -- the one that minimizes the dispersion of the error. The dispersion is well defined for the stable family and generalizes the Gaussian variance. 

Consider the same model~(\ref{eq:linmixgen}), an $a \in \mathbb{R}$ and consider a linear estimator of the form 
\begin{equation*}
\hat{X}(Y) = aY,
\end{equation*}
yielding an estimation error
\begin{equation*}
\hat{X}(Y) - X = (a_{21}a - a_{11})Z_1 + (a_{22}a-a_{12})Z_2,
\end{equation*}
which follows a \SaS distribution with scale parameter
\begin{equation}
\label{eq:errscale}
\gamma_e(a) =
\left[ |a_{21}a - a_{11}|^\alpha \gamma_{Z_1}^\alpha + |a_{22}a-a_{12}|^\alpha \gamma_{Z_2}^\alpha \right]^{\frac{1}{\alpha}} .
\end{equation}

Our goal is to find
\begin{equation}
\hat{a}
= \operatorname*{arg\,min}_{a \in \mathbb{R}}
\left[ |a_{21}a - a_{11}|^\alpha \gamma_{Z_1}^\alpha + |a_{22}a-a_{12}|^\alpha \gamma_{Z_2}^\alpha  \right].
\label{eq:minerror}
\end{equation}

\begin{theorem}[Dispersion-Optimal Linear Estimator]
\label{thm:optimal_a_unified}
Consider the linear estimator $\hat{X}(Y)=aY$. 
The coefficient $a$ minimizing the dispersion $\gamma_e$  of the estimation error --given by equation~\eqref{eq:minerror}-- is equal to

\begin{equation*}
\hat{a}=\frac{(|a_{21}|\gamma_{Z_1})^{\frac{\alpha}{\alpha-1}} \left[\frac{a_{11}}{a_{21}}\right] + (|a_{22}|\gamma_{Z_2})^{\frac{\alpha}{\alpha-1}}\left[\frac{a_{12}}{a_{22}}\right]}{(|a_{21}|\gamma_{Z_1})^{\frac{\alpha}{\alpha-1}} + (|a_{22}|\gamma_{Z_2})^{\frac{\alpha}{\alpha-1}}}  
\end{equation*}
if $\alpha>1$, and to
\begin{equation*}
\hat{a} =
\begin{cases}
\dfrac{a_{11}}{a_{21}}, & |a_{21}\gamma_{Z_1}|>|a_{22}\gamma_{Z_2}|, \\[2ex]
\dfrac{a_{12}}{a_{22}}, & |a_{21}\gamma_{Z_1}|<|a_{22}\gamma_{Z_2}|,\\
\text{any}, &  |a_{21}\gamma_{Z_1}|=|a_{22}\gamma_{Z_2}|
\end{cases}
\end{equation*}
whenever $\alpha\leq1$.
\end{theorem}

Note that when $|a_{21}\gamma_{Z_1}|=|a_{22}\gamma_{Z_2}|$, either value $\dfrac{a_{11}}{a_{21}}$ or $\dfrac{a_{12}}{a_{22}}$ achieves optimality so $\hat{a}$ can be arbitrarily chosen.

\begin{IEEEproof}
Letting  $k_1 \triangleq \frac{a_{11}}{a_{21}}$, $k_2 \triangleq \frac{a_{12}}{a_{22}}$, we can write $\gamma_e^\alpha(a)$ as follows
\[\gamma_e^\alpha(a) = |a_{21}|^\alpha |a - k_1|^\alpha \gamma_{Z_1}^\alpha + |a_{22}|^\alpha |a - k_2|^\alpha \gamma_{Z_2}^\alpha.\]

The first and second derivatives of the function $|x-b|^\alpha$ are defined everywhere except at $b$ and are equal to
\begin{align*}
    \frac{d}{dx}|x-b|^\alpha&=
    \begin{cases}
        \alpha (x-b)^{\alpha-1}, & x > b\\
        -\alpha (b-x)^{\alpha-1}, & x < b\\
    \end{cases}\\
    & = \alpha \, \sgn(x-b)|x-b|^{\alpha-1},
    \quad x \neq b 
    \end{align*}
    \begin{align*}
    \frac{d^2}{dx^2}|x-b|^\alpha & =
    \begin{cases}
        \alpha (\alpha - 1) (x-b)^{\alpha-2}, & x > b\\
        \alpha (\alpha - 1) (b-x)^{\alpha-2}, & x < b\\
    \end{cases}\\
    & = \alpha \, (\alpha -1 ) |x-b|^{\alpha-2}, \quad x \neq b.
\end{align*}

Determining the extremums of $\gamma_e^\alpha(a)$: For $a \neq k_1$ and $a \neq k_2$,
\begin{multline}
    \frac{d \gamma_e^\alpha}{d a}(a)
    =\alpha|a_{21}|^\alpha\sgn(a-k_1)|a-k_1|^{\alpha-1}\gamma_{Z_1}^\alpha \\
    +\alpha|a_{22}|^\alpha\sgn(a-k_2)|a-k_2|^{\alpha-1}\gamma_{Z_2}^\alpha
    \label{eq:dispderv}
\end{multline}
\begin{multline}
    \frac{d^2 \gamma_e^\alpha}{d a^2}(a)
    = \alpha(\alpha-1)|a_{21}|^\alpha|a-k_1|^{\alpha-2}\gamma_{Z_1}^\alpha \\
    + \alpha(\alpha-1)|a_{22}|^\alpha|a-k_2|^{\alpha-2}\gamma_{Z_2}^\alpha\label{eq:dispderv2}
\end{multline}

Let's assume first that $k_1 < k_2$.  
\begin{itemize}
    \item Case $\alpha \neq 1$
        \begin{enumerate}
            \item When $a < k_1$
            \begin{align*}
                \frac{d \gamma_e^\alpha}{d a}(a)
                &=-\alpha|a_{21}|^\alpha|a-k_1|^{\alpha-1}\gamma_{Z_1}^\alpha\\
                & \qquad \qquad -\alpha|a_{22}|^\alpha|a-k_2|^{\alpha-1}\gamma_{Z_2}^\alpha\\
                &\implies \frac{d \gamma_e^\alpha}{d a}(a)<0
            \end{align*}
            $\gamma_e(a)$ is decreasing, and $\arg \inf(\gamma_e)=k_1$ in the range $a \in (-\infty, k_1]$. \\
            
            \item When $k_1 \leq a \leq k_2$

            \begin{multline*}
                \frac{d \gamma_e^\alpha}{d a}(a)
                =\alpha|a_{21}|^\alpha|a-k_1|^{\alpha-1}\gamma_{Z_1}^\alpha\\
                -\alpha|a_{22}|^\alpha|a-k_2|^{\alpha-1}\gamma_{Z_2}^\alpha.
            \end{multline*}
            We want to find $\hat{a}$ such that
            \[\frac{d \gamma_e^\alpha}{d a}(\hat{a})=0,\]
            equivalent to
            \begin{align*}
                \alpha|a_{21}|^\alpha|\hat{a}-k_1|^{\alpha-1}\gamma_{Z_1}^\alpha
                &=\alpha|a_{22}|^\alpha|\hat{a}-k_2|^{\alpha-1}\gamma_{Z_2}^\alpha\\
                |a_{21}|^{\frac{\alpha}{\alpha-1}}(\hat{a}-k_1)\gamma_{Z_1}^{\frac{\alpha}{\alpha-1}}
                &=|a_{22}|^{\frac{\alpha}{\alpha-1}}(k_2-\hat{a})\gamma_{Z_2}^{\frac{\alpha}{\alpha-1}}
                \end{align*}
                \begin{align*}
                \left[(|a_{21}|\gamma_{Z_1})^{\frac{\alpha}{\alpha-1}}+(|a_{22}|\gamma_{Z_2})^{\frac{\alpha}{\alpha-1}}\right]\hat{a} =&\\\left[\frac{a_{11}}{a_{21}}\right]|a_{21}|^{\frac{\alpha}{\alpha-1}}&\gamma_{Z_1}^{\frac{\alpha}{\alpha-1}}\\+\left[\frac{a_{12}}{a_{22}}\right]|a_{22}|^{\frac{\alpha}{\alpha-1}}&\gamma_{Z_2}^{\frac{\alpha}{\alpha-1}},
                \end{align*}
                which implies
                \begin{multline*}
                \hat{a} = \frac{|a_{21}|^{\frac{1}{\alpha-1}}\gamma_{Z_1}^{\frac{\alpha}{\alpha-1}} \left[\frac{a_{11}}{a_{21}}\right]}{(|a_{21}|\gamma_{Z_1})^{\frac{\alpha}{\alpha-1}}+(|a_{22}|\gamma_{Z_2})^{\frac{\alpha}{\alpha-1}}}\\+\frac{|a_{22}|^{\frac{\alpha}{\alpha-1}}\gamma_{Z_2}^{\frac{\alpha}{\alpha-1}} \left[\frac{a_{12}}{a_{22}}\right]}{(|a_{21}|\gamma_{Z_1})^{\frac{\alpha}{\alpha-1}}+(|a_{22}|\gamma_{Z_2})^{\frac{\alpha}{\alpha-1}}}
                \end{multline*}
            The function $\gamma_e^\alpha(a)$ is convex whenever $\alpha > 1$, so $\hat{a}$ is a minimum for $\alpha>1$.

            When $\alpha<1$, $\hat{a}$ becomes a local maximum, and the value of $a$ that minimizes the optimization problem in this range will be either $k_1$ or $k_2$.
            
            \[\gamma_e^\alpha(k_1)=|a_{22}|^\alpha|k_1-k_2|^\alpha\gamma_{Z_2}^\alpha\]
            \[\gamma_e^\alpha(k_2)=|a_{21}|^\alpha|k_2-k_1|^\alpha\gamma_{Z_1}^\alpha\]
           and
            \begin{align*}
            \hat{a} =
            \begin{cases}
            \dfrac{a_{11}}{a_{21}}, 
            & |a_{21}|\gamma_{Z_1} > |a_{22}|\gamma_{Z_2}, \\[6pt]
            \dfrac{a_{12}}{a_{22}}, 
            & |a_{21}|\gamma_{Z_1} < |a_{22}|\gamma_{Z_2}, \\[6pt]
            a \in \left\{\dfrac{a_{11}}{a_{21}},\,\dfrac{a_{12}}{a_{22}}\right\},
            & |a_{21}|\gamma_{Z_1} = |a_{22}|\gamma_{Z_2}.
            \end{cases}
            \end{align*}

            \item When $a > \dfrac{a_{12}}{a_{22}}$
            \begin{align*}
                \frac{d \gamma_e^\alpha}{d a}(a)
                &=\alpha|a_{21}|^\alpha|a-k_1|^{\alpha-1}\gamma_{Z_1}^\alpha\\
                &\qquad \qquad +\alpha|a_{22}|^\alpha|a-k_2|^{\alpha-1}\gamma_{Z_2}^\alpha\\
                &\implies\frac{\partial\gamma_e^\alpha}{\partial a}(a)>0
            \end{align*}
            $\gamma_e(a)$ is increasing, and $\arg \inf(\gamma_e)=k_2$ in the range $a \in [k_2, \infty)$.
        \end{enumerate}
        
    \item Case $\alpha = 1$
        \begin{multline*}
            \frac{d \gamma_e}{d a}(a)
            =\alpha|a_{21}|\sgn(a-k_1)\gamma_{Z_1}\\
            +\alpha|a_{22}|\sgn(a-k_2)\gamma_{Z_2}
        \end{multline*}
        \begin{enumerate}
            \item When $a < k_1$
            \[\frac{d \gamma_e}{d a}(a)=-\alpha|a_{21}|\gamma_{Z_1}-\alpha|a_{22}|\gamma_{Z_2} < 0\]
            $\gamma_e(a)$ is decreasing, and $\arg \inf(\gamma_e) = k_1$ in this range.\\
            
            \item When $k_1 \leq a \leq k_2$
            \[\frac{d \gamma_e}{d a}(a)=\alpha|a_{21}|\gamma_{Z_1}-\alpha|a_{22}|\gamma_{Z_2}\]
            Based on the sign of $|a_{21}|\gamma_{Z_1}-|a_{22}|\gamma_{Z_2}$, we obtain
            \begin{equation*}
            \hat{a} =
            \begin{cases}
            \dfrac{a_{11}}{a_{21}}, 
            & |a_{21}|\gamma_{Z_1} > |a_{22}|\gamma_{Z_2}, \\[6pt]
            \dfrac{a_{12}}{a_{22}}, 
            & |a_{21}|\gamma_{Z_1} < |a_{22}|\gamma_{Z_2}, \\[6pt]
            a \in \left\{ \dfrac{a_{11}}{a_{21}},\,\dfrac{a_{12}}{a_{22}}\right\},
            & |a_{21}|\gamma_{Z_1} = |a_{22}|\gamma_{Z_2}.
            \end{cases}
            \end{equation*}
            
            \item When $a > k_2$
            \[\frac{d \gamma_e}{d a}(a)=\alpha|a_{21}|\gamma_{Z_1}+\alpha|a_{22}|\gamma_{Z_2}>0\]
            $\gamma_e(a)$ is increasing, and $\arg \inf(\gamma_e)=k_2$ in this range.
        \end{enumerate}

\end{itemize}

If on the other hand $k_1 > k_2$, the analysis is conducted in an identical fashion with the sub-indices 1 and 2 swapped and the results are identical.

\end{IEEEproof}

\section{Sub-Gaussian}


\subsection{The conditional law}
Let $A \sim \mathcal{S}\left(\frac{\alpha}{2},1,\cos^{\frac{2}{\alpha}}\left(\frac{\pi \alpha}{4}\right),0\right)$ be a positive $\alpha/2$-stable random variable, $0 < \alpha < 2$, and let $(G_1, G_2) \sim \mathcal{N}(\mathbf{0}, \Sigma)$, \(\Sigma =  \begin{pmatrix}
    \sigma^2_{G_{1}} & \rho \sigma_{G_1} \sigma_{G_2} \\ 
    \rho \sigma_{G_1} \sigma_{G_2} & \sigma^2_{G_2}
    \end{pmatrix}\),  be independent of $A$. Define
\[
(X, Y) = \sqrt{A}\,(G_1, G_2),
\]
so that $(X,Y)$ is an elliptically contoured $\alpha$-stable vector. This model is well-known under the name sub-Gaussian \SaS vector~\cite{sam1994}. It finds application in modeling high dimensional features~\cite{beck2003,Adomaityte2023ClassificationOH}. Other examples can be found in~\cite{wainwright1999} and~\cite{Adomaityte2023HighdimensionalRR}. We present next two useful technical lemmas. 

\begin{lemma}[The Conditional Representation]
\label{lem:condrep}
Let $(X,Y) = \sqrt{A}\,(G_1, G_2)$ be a sub-Gaussian \SaS vector with underlying Gaussian vector $(G_1, G_2) \sim \mathcal{N}(\mathbf{0}, \Sigma)$.
    Conditionally on $Y$, the random variable $X$ admits the stochastic representation
\[
X \mid Y \;\overset{d}{=}\; \rho\,\frac{\sigma_{G_1}}{\sigma_{G_2}}\,Y + \sqrt{A_Y}\,Z,
\]
where $A_Y \;=\; A \mid Y$, $Z \sim \mathcal{N}\big(0, \sigma_{G_1}^2(1-\rho^2)\big)$, and $A_Y$ and $Z$ are conditionally independent given $Y$.
\end{lemma}

\begin{IEEEproof}
Conditionally on $A=a$, we have
\[
(X, Y) \mid (A=a) \sim \mathcal{N}(0, a\Sigma),
\]
which implies
\begin{equation}
X \mid (Y=y, A=a) = \beta y + \sqrt{a}\,\sigma_{G_1}\sqrt{1-\rho^2}\,\tilde{Z}, \label{eq:condYA}
\end{equation}
where $\beta \triangleq \rho \frac{\sigma_{G_1}}{\sigma_{G_2}}$, $\tilde{Z} \sim \mathcal{N}(0, 1)$ and
\(
\tilde{Z} \;\perp\!\!\!\perp\; Y \mid A.
\)
Marginalizing over $A$, we obtain
\[
X \mid (Y=y) \;\overset{d}{=}\; \beta y + \sqrt{A_y}\,Z,
\]
where $A_y \triangleq A \mid (Y=y)$, $Z \sim \mathcal{N}\big(0, \sigma_{G_1}^2(1-\rho^2)\big)$, and
\(
A_y \;\perp\!\!\!\perp\; Z \mid Y=y
\).
\end{IEEEproof}

\begin{lemma}
[Integrability of the residual term]
\label{lem:integrability}
Let $(X,Y) = \sqrt{A}\,(G_1, G_2)$ be a sub-Gaussian \SaS vector with $(G_1,G_2) \sim \mathcal{N}({\bf 0}, \Sigma)$. Let $X \mid Y$ admit the stochastic representation
\[
X \mid Y \;\overset{d}{=}\; \rho \frac{\sigma_{G_1}}{\sigma_{G_2}} Y + \sqrt{A_Y}\,Z,
\]
as in Lemma~\ref{lem:condrep}, where $Z \sim \mathcal{N}\big(0, \sigma_{G_1}^2(1-\rho^2)\big)$ and $A_Y = A \mid Y$. Then, the residual term $\sqrt{A_Y} Z$ is integrable:
\[
\mathbb{E}[\,|\sqrt{A_Y} Z| \;\mid\; Y] < \infty \quad \text{almost surely}.
\]
\end{lemma}
\begin{IEEEproof}
Fix $Y=y$. We have
\[
\sqrt{A_Y} Z \;\big|\; (Y=y) \; =\; \sqrt{A_y}\, Z,
\]
where $A_y \triangleq A \mid (Y=y)$ and $Z \perp A_y$. By independence, we write
\[
\mathbb{E}[\,|\sqrt{A_y} Z|\,] = \mathbb{E}[\sqrt{A_y}] \, \mathbb{E}[\,|Z|\,].
\]
Since $Z \sim \mathcal{N}(0, \sigma_{G_1}^2(1-\rho^2))$ has a finite absolute first moment,
\[
\mathbb{E}[\,|Z|\,] = \sigma_{G_1} \sqrt{2(1-\rho^2)/\pi} < \infty,
\]
it remains to show that $\mathbb{E}[\sqrt{A_y}] < \infty$. Using the conditional distribution of $A$ given $Y=y$, we have
\[
p_{A\mid Y=y}(a) \propto a^{-1/2} \exp\Big(-\frac{y^2}{2 a \sigma_{G_1}^2}\Big) p_A(a),
\]
where $p_A(a)$ is the density of $A$, continuous on $\mathbb{R}^+$~\cite{kolmo1968}. For large values of $a > 0$, the exponential term is bounded and $p_A(a) = \Theta(a^{-1-\alpha/2})$~\cite{kolmo1968}, which implies that $p_{A\mid Y=y}(a) = \Theta(a^{-3/2-\alpha/2})$. Then
\[
\int_0^\infty a^{1/2} p_{A\mid Y=y}(a)\,da < \infty \quad \text{for all } \alpha > 0,
\]
so $\mathbb{E}[\sqrt{A_y}] < \infty$. Combining these facts,
\[
\mathbb{E}[\,|\sqrt{A_Y} Z| \;\mid\; Y=y] = \mathbb{E}[\sqrt{A_y}]\,\mathbb{E}[|Z|] < \infty,
\]
which proves integrability almost surely.
\end{IEEEproof}
\begin{corollary}[Conditional Mean is Linear in $Y$]
\label{corr:condmean}
Whenever $(X,Y) = \sqrt{A}\,(G_1, G_2)$ is a sub-Gaussian \SaS vector, the conditional expectation of $X$ given $Y=y$ exists and is linear:
\[
\mathbb{E}[X \mid Y=y] \;=\; \rho \frac{\sigma_{G_1}}{\sigma_{G_2}}\,y.
\]
\end{corollary}

\begin{IEEEproof}
By Lemma~\ref{lem:condrep}, for every fixed $Y=y$ we have
\[
X \mid (Y=y) = \beta y + \sqrt{A_y}\,Z,
\]
for $\beta = \rho \frac{\sigma_{G_1}}{\sigma_{G_2}}$. Lemma~\ref{lem:integrability} shows that the residual term $\sqrt{A_y}Z$ is integrable conditional on $Y=y$, i.e.,
\[
\mathbb{E}\left[\left|\sqrt{A_y}Z\right|\right] < \infty.
\]
Therefore, the conditional expectation of $X$ exists and 
\begin{align*}
\mathbb{E}[X \mid Y=y] 
&= \mathbb{E}[\beta y + \sqrt{A_y}Z \mid Y=y] \\ 
&= \beta y + \mathbb{E}[\sqrt{A_y}Z \mid Y=y].
\end{align*}

Since $A_y$ and $Z$ are conditionally independent and $\mathbb{E}[Z]=0$, we obtain
\[
\mathbb{E}[\sqrt{A_y}Z \mid Y=y]
= \mathbb{E}[\sqrt{A_y}]\cdot\mathbb{E}[Z] = 0.
\]
Thus,
\[
\mathbb{E}[X \mid Y=y] = \beta y.
\]
\end{IEEEproof}

\subsection{The Dispersion-Optimal Linear Estimator}
\begin{theorem}[Optimal Linear Coefficient under a Dispersion Constraint (sub-Gaussian version)]
Let $(X,Y) = \sqrt{A}\,(G_1, G_2)$ be a sub-Gaussian \SaS vector with $(G_1,G_2) \sim \mathcal{N}({\bf 0}, \Sigma)$. Consider the linear estimator 
\begin{equation*}
\hat{X}(Y) = b\,Y,
\end{equation*}
yielding an estimation error
\begin{equation*}
E = \hat{X}(Y) - X = \sqrt{A}(b\,G_2-G_1),
\end{equation*}
The coefficient $b$ minimizing the scale $\gamma_E$ of the estimation error is \[\hat{b} = \beta = \rho\frac{\sigma_{G1}}{\sigma_{G_2}}\]
\end{theorem}

\begin{IEEEproof}
Let $G = b\,G_2 - G_1$. Since $(G_1,G_2)$ is jointly Gaussian, we have 
$G \sim \mathcal{N}(0,\sigma_G^2(b))$ where $\sigma_G^2(b) \triangleq \text{Var}(b\,G_2 - G_1)$. 
Conditionally on $A = a$, 
\(
E | (A = a) \sim \mathcal{N}(0, a\,\sigma_G^2(b)).
\)
Hence, the conditional characteristic function is
\[
\phi_{E\mid A}(t) = \exp\!\left(-\tfrac12 A\sigma_G^2(b) t^2\right).
\]
By the law of total expectation, the characteristic function of $E$ becomes
\[
\phi_E(t)
= \E{\phi_{E\mid A}(t)}
= \E{\exp(-\tfrac12 A\sigma_G^2(b) t^2)}.
\]

Since $A$ is a positive $\alpha/2$-stable variable with $\gamma_A = \cos^{\frac{2}{\alpha}}\left(\frac{\pi \alpha}{4}\right)$, its Laplace transform is equal to~\cite[Sec.~2.5]{sam1994}
\[
\E{e^{-uA}} = \exp\!\left(-\, |\gamma_A u|^{\alpha/2}\right). 
\]
We may substitute $u = \tfrac12 \sigma_G^2(b) t^2$ to obtain
\[
\phi_E(t)
= \exp\!\left(-\left(\tfrac12 \sigma_G^2(b) \gamma_A\right)^{\alpha/2} |t|^\alpha\right),
\]
which represents the characteristic function of a \SaS variable with dispersion
\[
\gamma(b) = \left(\tfrac12\sigma_G^2(b)\gamma_A\right)^{\alpha/2}.
\]

Thus, minimizing $\gamma_E(b)$
is equivalent to minimizing 
\[\sigma_G^2(b) = \text{Var}(b\,G_2 - G_1) = b^2 \sigma^2_{G_2} + \sigma^2_{G_1} - 2\,b\,\rho\,\sigma_{G_1} \sigma_{G_2},\] 
which yields
\[
b^* = \beta = \rho\frac{\sigma_{G_1}}{\sigma_{G_2}}.
\]
\end{IEEEproof}

\subsection{A Linear Maximum A Posteriori Estimator}
\begin{theorem}[MAP optimality]
Let $(X,Y) = \sqrt{A}\,(G_1, G_2)$ be a sub-Gaussian \SaS vector with $(G_1,G_2) \sim \mathcal{N}({\bf 0}, \Sigma)$. 
Then the maximum a posteriori (MAP) estimator of $X$ given $Y$ is
\[
X^*(Y)=\rho\,\frac{\sigma_{G_1}}{\sigma_{G_2}}\,Y.
\]
\end{theorem}

\begin{IEEEproof}
Using equation~\eqref{eq:condYA}, the conditional density of $X$ given $(Y=y,A=a)$ is therefore
\[
p_{X \mid (Y,A)}(x \mid (y,a))
= \frac{1}{\sqrt{2\pi a\tau^2}}
\exp\!\left(-\frac{(x-\beta y)^2}{2a\tau^2}\right),
\]
where $\tau^2=\sigma_{G_1}^2(1-\rho^2)$. 
Marginalizing over $A$, we obtain
\begin{eqnarray*}
p_{X \mid Y}(x \mid y)
&=& \int_0^\infty p_{X \mid (Y,A)}(x \mid (y,a))\,p(a)\,da\\
&\leq& \int_0^\infty \frac{1}{\sqrt{2\pi a\tau^2}}\,p(a)\,da < \infty,
\end{eqnarray*}
with equality if and only if $x = \beta y$. The finiteness of the integral term is due to the fact that $p_A(a)$ is continuous on $\mathbb{R}^+$ and is $\Theta(a^{-1-\alpha/2})$ for large values of $a$~\cite{kolmo1968}. Consequently, for every $y$, the posterior density $p_{X \mid Y}(x \mid y)$ is maximized at
\[
x^*(y)= \beta y = \rho \frac{\sigma_{G_1}}{\sigma_{G_2}}y.
\]
\end{IEEEproof}






\bibliographystyle{IEEEtran}
\bibliography{references}









\end{document}